\documentclass[11pt]{article}

\usepackage[margin=1in]{geometry}
\usepackage[T1]{fontenc}
\usepackage{lmodern}
\usepackage{microtype}
\usepackage{amsmath,amssymb}
\usepackage{booktabs}
\usepackage{tabularx}
\usepackage{enumitem}
\usepackage{xcolor}
\usepackage[hidelinks]{hyperref}
\usepackage{url}

\newcommand{\repo}{\url{https://github.com/henryzhangpku/autonomous-quant-researcher}}

\newcommand{\paperauthor}{Henry Zhang\\\paperaffiliation\\\texttt{github.com/henryzhangpku}}
\newcommand{\aboutauthor}{%
\paragraph{About the author.} Henry Zhang is a quantitative engineer with
fifteen years building research platforms and production trading systems,
including eight years at BlackRock and two at PIMCO. The research engine used here
is open source at
\url{https://github.com/henryzhangpku/autonomous-quant-researcher}; more
research and notes at \url{https://henryzhang.substack.com} and
\url{https://www.officialhenryzhang.com}.}

\title{Hard to Fool: A Bounded LLM Research Loop\\for Quantitative Finance, and What Two Dozen\\Campaigns Found}
\author{\paperauthor}
\date{November 2026}

\begin{document}
\maketitle

\begin{abstract}
Agentic research loops that let a language model propose a change, run a fixed
experiment and keep whatever improves the metric work well in machine learning,
where a lower validation loss is usually a better model. In trading research
the same loop is a machine for producing false discoveries: a backtest that
``improves the metric'' is, by default, overfit, look-ahead, a
multiple-comparisons artifact, cost-free, or market drift wearing a strategy.
We describe Autonomous Quant Researcher, an open-source loop built on the
opposite premise: the model is allowed to propose, and only trusted code is
allowed to decide. Proposals are strict declarative JSON over a fixed
vocabulary, so there is no model-written code to sandbox. Evaluation runs on
immutable chronological stores separated into discovery, validation and a final
holdout that can be opened exactly once, after a durable receipt is written.
Acceptance is a conjunction of absolute gates (positive edge after costs,
positive net P\&L, a positive weekly-cluster bootstrap lower bound, survival at
elevated costs, per-year stability and sample floors), never a single score.
Duplicates and near-duplicates are refused before they spend a trial, and every
formulation, refusal, crash and result is recorded in a hash-chained,
append-only ledger. We report every campaign the system has run to date,
across SPY and QQQ zero-days-to-expiry (0DTE) options priced from actual traded
option bars, overnight SPX verticals, daily ETF rules, the premises behind
popular technical-analysis scripts, Kalshi crypto binaries, earnings options,
off-exchange volume and a pre-registered alternative-data campaign. Most
hypotheses fail, and we report them. Two results survive their own
pre-registered bars. The clearest example of the loop working is a failure: a
language-model survivor that passed discovery and validation with Sharpe
ratios near 3 and fell to 0.26 on its single holdout use. We argue that
the scientific value of such a loop lies less in what it finds than in how
hard it is to make it report something that is not there.
\end{abstract}

\section{Introduction}

Two open projects released in 2026 made a simple observation about agents: the
agent is the cheap part, and the loop around it is the product. Karpathy's
\texttt{autoresearch}~\cite{karpathy2026autoresearch} lets an agent edit a
single training script, run a fixed five-minute job, keep the change if the
validation metric improved and discard it otherwise, all night. Greyling's Loop
Engineering~\cite{greyling2026loop} turns the loop itself into a first-class
artifact: a persistent contract, explicit budgets, deterministic gates, an
audit trail and maker/checker separation.

Finance is a domain where that contract breaks in a specific way. The problem
of data snooping in asset pricing is old~\cite{lo1990data,white2000reality},
with formal corrections for searching many rules at once~\cite{hansen2005test,romano2005stepwise},
and the evidence that most published return predictors do not survive
correction for the number of tests tried is now
extensive~\cite{harvey2016cross,hou2020replicating}. Backtest overfitting is
easy to produce and hard to detect after the
fact~\cite{bailey2014pseudo,bailey2014deflated,bailey2017probability}, and published
predictability decays once it is known~\cite{mclean2016does}. A language model that
proposes thousands of hypotheses per night does not repair any of this. It
industrializes it.

Recent LLM-agent systems for trading and alpha mining, such as
TradingAgents~\cite{xiao2024tradingagents}, FinMem~\cite{yu2023finmem},
AlphaAgent~\cite{tang2025alphaagent},
R\&D-Agent-Quant~\cite{li2025rdagentquant}, tree-search factor
mining~\cite{shi2025alphajungle} and code-based alpha
evolution~\cite{liu2025cognitive}, and role-based agents for equity
portfolio construction~\cite{zhao2025alphaagents}, focus on making the
proposer better: richer roles, memory, reflection, factor generation and
reinforcement. Others ask what a language model can extract in the first
place, for example whether it tracks evolving signals in corporate
disclosures~\cite{choi2025texttoalpha}. Work on the evaluation side is newer. Deep et
al.~\cite{deep2025walkforward} validate interpretable hypotheses walk-forward
across many out-of-sample periods with costs; Li et
al.~\cite{li2025timetravel} argue that historical backtests let a language
model ``time travel'' through outcomes it has already seen, and benchmark live
instead; and a scoping review by Yao, Zheng and Li~\cite{yao2026execution}
finds execution realism weakly standardized across LLM trading studies, which
limits comparison and reproduction. Closest to our aim,
Gen\c{c}ay~\cite{gencay2026honest} builds leakage safety and search-aware
deflation into a strategy-discovery system and finds that no LLM-discovered
strategy survives certification at the agent's own trial count; a planted
look-ahead oracle with a Sharpe ratio of 35 passes both the deflated Sharpe
ratio and the probability of backtest overfitting, so statistical correction
cannot stand in for structural leakage control. Chen and
Pu~\cite{chen2026nowcasting} remove look-ahead at inference by logging an
agentic model's daily stock scores in real time. Evidence from production
points the same way: across two live fleets of LLM trading agents, Barton et
al.~\cite{barton2026production} find that the operating layer (sizing
constraints, rendered choices, order mechanics) shapes behaviour more than the
choice of model, and that neither fleet shows a directional edge. We share
Gen\c{c}ay's verdict and differ in the instrument: hypotheses are declared as
data and pre-registered, the holdout can be spent once, and every campaign,
across options, microstructure, prediction markets and equities, is reported
with its rejections. We focus on that other half of the loop. Our contribution is a
reference design, and an open implementation, for the part that decides
whether a proposal is true, together with a complete report of what it
decided.

Concretely, this paper contributes:
\begin{enumerate}[leftmargin=*,itemsep=2pt]
\item A \textbf{declarative hypothesis boundary}: the model emits data, not
code, so the attack surface and the space of searchable hypotheses are both
fixed in advance (Section~\ref{sec:boundary}).
\item A \textbf{staged evidence lifecycle} with a one-use holdout enforced by a
durable receipt, admission control that refuses duplicates before they spend a
trial, and acceptance by a conjunction of absolute gates
(Section~\ref{sec:lifecycle}).
\item A \textbf{tamper-evident ledger} and portable findings that carry their
own provenance and authority (Section~\ref{sec:ledger}).
\item A \textbf{complete campaign report}, including every rejection (Section~\ref{sec:results}).
\item A list of \textbf{failure modes that the gates did not catch} the first
time, and the change each one forced (Section~\ref{sec:lessons}).
\end{enumerate}

The engine is open source~\cite{zhang2026engine}, with a reference mission
and a public pre-registration whose ledger can be verified independently.
Market-data files are not redistributed.

\section{Design}

\subsection{The declarative hypothesis boundary}\label{sec:boundary}

In the current schema (version~3) a proposal has exactly two fields,
\texttt{hypothesis} and \texttt{rationale}. The hypothesis contains a schema
version, an entry minute on the five-minute regular-session grid, a bounded
Boolean signal expression over approved causal observations (return from the
open, return between two earlier times, first touch before entry) and exactly
one approved defined-risk options structure (a call debit spread or a put
credit spread) with bounded dimensions. Trusted functions interpret the
declaration. A response with extra fields, invalid JSON, unsupported
vocabulary, malformed timing, excessive nesting or out-of-range parameters is
rejected before any scientific evaluation.

This boundary replaces an earlier design, preserved in the repository, in
which the model returned an executable Python candidate. That design needed a
sandbox (credential stripping, import and capability denial, a content-hashed
evaluator) and it still left the candidate free to express things the
evaluator did not anticipate. Removing the executable field removes the
problem rather than containing it: there is nothing to sandbox, and every
admissible hypothesis is, by construction, one the evaluator knows how to
price.

\subsection{Staged evidence and the one-use holdout}\label{sec:lifecycle}

Stage stores are materialized together from one historical source, but each
access capability resolves only its own stage, and every store carries a
content hash and a manifest hash. Each campaign freezes its own windows. In
the reference SPY mission on daily and modeled data, discovery is 2016--2021,
validation 2022--2024 and the final holdout 2025 onward. The 0DTE campaigns
priced from traded option bars start when those bars do: discovery is
2024-01-19 to 2025-08-05, validation 2025-08-06 to 2026-02-05 and the holdout
2026-02-06 to 2026-08-10.

\paragraph{Freeze before the first result.} The mission, policy, vocabulary,
interpreter and evaluator versions, stage manifests, base and elevated costs,
trial budget, diversity rules and gates are hashed into a campaign contract
before the first proposal is evaluated. Changing any of them creates a new
campaign; a non-terminal campaign cannot continue under changed semantics.

\paragraph{Admission.} Formulations and scientific trials are counted
separately. Invalid documents are recorded but do not spend a trial. Semantic
duplicates and near-duplicates (one numeric knob turned on the same idea) are
refused, and a family quota admits at most two candidates per descriptor
family. A 50-trial budget therefore buys 50 distinct ideas, and the trial
count reported with any result is the number that was actually tried.

\paragraph{Shortlist.} After the discovery budget closes, only candidates that
passed every discovery gate are eligible. The coordinator ranks them by their
weakest gate component, prefers distinct clusters, freezes at most three and
hashes the shortlist. Validation evaluates only that set and divides the 5\%
lower-tail probability by the shortlist size.

\paragraph{The holdout.} The ordinary worker stops at \texttt{holdout\_ready}
and cannot open the final stage. Holdout access is resolved only after an
irreversible receipt has been written durably; a crash after the receipt
still consumes the holdout and cannot be retried. Budget exhaustion is a
pause, never a success.

\subsection{Gates, not scores}

Every stage must pass every applicable gate:
\begin{itemize}[leftmargin=*,itemsep=1pt]
\item absolute edge after base costs $>0$;
\item base-cost mean net P\&L $>0$;
\item weekly-cluster bootstrap lower bound on net P\&L $>0$ (2{,}000
deterministic resamples, clustered by ISO week; cf.\ block
bootstrap~\cite{kunsch1989jackknife});
\item mean net P\&L at elevated costs $>0$;
\item finite metrics and priced rows;
\item sample floors. In the reference SPY mission these are 120 trades and 40
effective weeks in discovery; 60 and 24 in validation, with evidence in each
of 2022, 2023 and 2024, at least 15 trades and 8 weeks per year, and no
negative year; and 30 and 12 in the holdout.
\end{itemize}
Floors scale with a campaign's windows and are frozen in its contract. The
traded-bar 0DTE campaigns, with shorter windows, require 40, 15 and 15 trades.
A ranking score, the minimum of absolute edge, the bootstrap bound,
elevated-cost P\&L and the worst year, chooses \emph{among} candidates that
passed; it cannot admit one. A control-relative metric is computed but is
diagnostic only, so a weak control cannot rescue a candidate.

Modeled option prices are not market prices. Actual OPRA evidence is a
separate typed gate with its own coverage floors and states
(\texttt{unavailable}, \texttt{insufficient\_coverage},
\texttt{failed\_evidence}, \texttt{passed}); replay consumes only a hash-verified
quote artifact with no fallback price source.

\subsection{Ledger, provenance and authority}\label{sec:ledger}

Every event (formulation, admission decision, trial, crash, stage transition,
holdout receipt) is appended to a ledger in which each record carries the hash
of its predecessor. Editing, removing or reordering an event makes the campaign
fail closed. A surviving finding is exported as portable JSON built only from
frozen aggregate evidence: contract and manifest hashes, per-stage evidence,
limitations and invalidation conditions, and an immutable authority stamp
(\texttt{research\_only}, \texttt{execution\_authorized: false}). Whatever
consumes a finding downstream knows exactly what it is holding.

The same machinery supports pre-registration for campaigns whose data does not
yet exist: a specification and the code that will compute its verdict are
bound by SHA-256 into the ledger, typed data requirements refuse to run until
met, and each campaign is judged once, with the receipt written before any data
is read~\cite{nosek2018preregistration}.

\subsection{Model independence}

The evaluator is deterministic CPU code; only the proposer needs a model. Any
OpenAI-compatible endpoint can propose. The reference configuration is a
locally served 7B-parameter coding model. Because the evaluator does not
depend on who proposes, changing the model cannot move a metric, only the
candidates.

\section{Results}\label{sec:results}

Table~\ref{tab:campaigns} lists every campaign recorded in the repository as of
this writing. Each campaign has a record holding its pre-registration (where one exists),
result tables and limitations.
Some campaigns predate schema~3 and ran through the executable-candidate loop
or a fixed sweep harness under the same staged lifecycle, with the splits,
costs and floors frozen in each campaign's own files; one (Kalshi maker side)
is explicitly labelled hypothesis generation, because its data had already
been read. We report all of them.

\begin{table}[!htbp]
\centering
\small
\begin{tabularx}{\textwidth}{@{}>{\raggedright}p{3.9cm}>{\raggedright}X>{\raggedright\arraybackslash}X@{}}
\toprule
Campaign & Question and scale & Verdict \\
\midrule
\multicolumn{3}{@{}l}{\emph{0DTE and overnight index options (actual traded option bars unless noted)}}\\
spy-odte-credit & SPY 0DTE put credit, w5/o5, 629 sessions & passes both stages at 0.5\% cost; fails drift-neutral bound \\
\quad geometry cross & 5 pre-registered geometries & 0 of 5; best (w5/o2) 7 of 8 conditions \\
\quad QQQ replication & one named cell, 635 sessions & 7 of 8; same failing condition \\
\quad calm conditioning & prior-day stress filter & refuted; mechanism inverted \\
odte-intervention & 120 management rules vs.\ hold & 0 supported \\
qqq-0dte-holdout & LLM survivor, holdout spent once & \textbf{rejected}: Sharpe 2.94 $\to$ 0.26 \\
side-asymmetry & post-hoc call/put asymmetry, 8 new markets & replication refuted \\
overnight-spx & 40 LLM trials, modeled SPX verticals & 0 passed discovery \\
\quad entry sweeps & pre-registered grids; w20/o15 promoted & holdout spent once, \textbf{rejected} \\
\quad 1DTE horizon & 228 cells, one full validation pass & rejected as drift (56\%) \\
odte-call-side, friday-odte-names & single-name 0DTE & not promoted \\
\midrule
\multicolumn{3}{@{}l}{\emph{Daily bars and technical analysis}}\\
bars-rules & 232 single-rule cells, 8 ETFs & 4 discovery passers, 0 promotions \\
ta-premises & Fibonacci, channels, MA ribbon, support & all fail validation \\
ta-indicators & 8 indicator claims & 0 cleared discovery \\
smartfit-channels & 5 pivot-channel claims & 0 cleared discovery \\
hmm-regime & HMM regime system, SPY daily, BTC hourly & rejected on both \\
idio-gap & idiosyncratic gap continuation, 1{,}945 events & refused; crash-day reversion found \\
earnings-vol, pead & earnings straddles, wings, drift & no edge \\
\midrule
\multicolumn{3}{@{}l}{\emph{Crypto and prediction markets}}\\
kalshi-btc-hourly & 138 conditional hypotheses & 0 cleared; one look-ahead artifact \\
kalshi-btc-15m & 15-minute binaries & not supported; fails replication \\
kalshi-maker-side & maker fills vs.\ adverse selection & no (hypothesis generation) \\
btc-sessions, btc-kalshi-windows & weekend and overnight BTC & 0 promotions \\
flow-crypto-etf-weekend & Friday ETF options flow, 84 cells & 0 cleared Bonferroni \\
\midrule
\multicolumn{3}{@{}l}{\emph{Cross-section and alternative data}}\\
dpi-offexchange & off-exchange share vs.\ forward returns & \textbf{clears pre-registered bar} \\
gpu-leads-revisions & GPU rental prices $\to$ revisions & primary waiting; proxy v3 fails \\
\bottomrule
\end{tabularx}
\caption{Every campaign run to date. ``w5/o2'' denotes a spread 5 points wide with the short strike 2
points out of the money. Costs and gates are those frozen in each campaign's
own files.}
\label{tab:campaigns}
\end{table}

\subsection{A language-model survivor meets its holdout}

The most instructive single result is a QQQ 0DTE campaign in which the
language-model proposer, in its thirteenth run, produced a rule that sold the
near-money put credit spread on Mondays and Tuesdays when the credit exceeded
10\% of width. On the traded-bar windows above, and under that campaign's
own floors (40, 15 and 15 trades), it passed every discovery and validation
gate (Table~\ref{tab:holdout}). On explicit instruction the holdout was spent once.
Eight of nine holdout gates passed; the rule stayed net positive, survived
doubled costs and the removal of its best day. But expectancy fell by about
90\%, the Sharpe ratio by an order of magnitude, and one half of the holdout
window was net negative. The rule was rejected, and the holdout window is now
permanently marked as spent for this candidate.

\begin{table}[!htbp]
\centering
\small
\begin{tabular}{@{}lrrr@{}}
\toprule
 & Discovery & Validation & Holdout \\
\midrule
Net per trade (\% of width) & $+5.69$ & $+4.82$ & $+0.52$ \\
Net Sharpe & 2.94 & 3.34 & 0.26 \\
Trades & 221 & 43 & 72 \\
Status & passed & passed & rejected \\
\bottomrule
\end{tabular}
\caption{The QQQ language-model survivor across its three stages.}
\label{tab:holdout}
\end{table}

Two points matter for the design. First, the in-sample numbers were not
marginal; a loop that accepts on a validation Sharpe above 3 would have
promoted this rule. Second, the frozen specification for that campaign had
been written to a temporary path and was not on disk when the holdout was
requested. Rather than evaluate under a guessed specification, it was
reconstructed and checked against the two already-spent stages (discovery
Sharpe matched the recorded value to six decimals, and the dataset hash
matched the staging manifest) before the holdout was read. The campaign
contract is now persisted with the ledger, so this cannot recur.

\subsection{The 0DTE put-credit premium: real, and not enough}

The unconditional SPY 0DTE put credit spread (width 5, short strike 5 points
out of the money, 10:00~ET entry, held to settlement) passed every discovery
and validation gate at a round-trip cost of 0.5\% of width, with $+55.7$ and
$+71.6$ basis points of width per session respectively, and a gross win rate
of 95.2\%. Three follow-ups, each pre-registered
as a new contract, then tested whether this was a strategy or a byproduct.

\emph{Drift.} Regressing each session's P\&L on SPY's own return over the
same window attributes 13--30\% of the payoff to market direction. The
drift-neutral residual is positive in both splits ($+74$ and $+100$ bps of
width gross) but its weekly bootstrap lower bound is marginally negative in
both. By the rule written before the attribution ran, the structure is not
promoted.

\emph{Geometry.} Across five pre-registered geometries the premium is
monotone in moneyness: offset 2 beats 5 beats 10, and offset 10 is negative.
The best cell, w5/o2, passes every condition except one, the discovery raw
bootstrap bound, which misses by 1.35 bps of width. A pre-registered QQQ
replication of the same cell in percentage terms fails exactly the same
condition at a similar magnitude. The worst discovery weeks on both tapes are
the same calendar shocks (April 2024, August--September 2024, March--April
2025).

\emph{Conditioning.} A filter that skips entry after a prior-day move of at
least 1\% (chosen by convention, not fit) makes every bootstrap worse. The
losses come from the first day of a shock, which follows a calm day and is not
forecastable from prior-day information; by the next day volatility has
repriced and selling pays best.

The honest summary is that the near-money 0DTE put premium is real, has the
same shape on two underlyings, is mostly not drift, and earns its return by
absorbing unforecastable first-day shocks. The pre-registered bar, which asks
for a positive fifth percentile of weekly means across a window containing
three such shocks, is not met, and the bar was not moved. The breakeven entry
cost for full gate compliance lies between 0.5\% and 1.0\% of width, which
turns an open question about the strategy into a measurable question about
execution.

\subsection{What the TA and daily-rule campaigns show}

In the 232-cell daily-rule grid, all four discovery passers used the same
feature in both directions: long after a 20-day decline and long after a
20-day advance. A condition whose opposite passes equally carries no
information; what passed was the long side in a window where holding these
ETFs paid. The gates caught sample and stability problems but could not by
themselves detect that the whole discovery window pointed one way. All three
shortlisted cells failed every performance gate in validation. In the technical-analysis campaigns, one premise (a
regression channel) passed all nine discovery gates and then lost 22 bps per
session in validation; every premise also underperformed simply being long.

\subsection{What survived}

Two results clear their own pre-registered bars and are the subject of
separate papers. The share of a stock's volume executed off-exchange, built
from public FINRA data, has a negative information coefficient at 1, 5 and 20
days ($t$ between $-3.1$ and $-5.8$), negative in every calendar year
2021--2026, concentrated in the top decile, and robust to size and short-term
reversal; the vendor-published ``dark pool index'' predicts nothing. In the idiosyncratic-gap campaign the
pre-registered continuation hypothesis was refused, while reversion of the
day's losers on market-wide crash days survived every gate. We report both as research findings with their
limitations, not as strategies.

\section{Failure modes the gates missed, and the change each forced}\label{sec:lessons}

\begin{description}[leftmargin=*,itemsep=3pt]
\item[Drift as alpha.] An unconditional long-delta credit spread collects the
market's upward drift whether or not anything is mispriced. An overnight SPX
1DTE at-the-money call debit spread passed every validation gate with a weekly
Sharpe ratio of 2.27; 56\% of its validation payoff was the underlying's own
move, and the bootstrap bound on the residual was negative. A pre-registered
drift control (regress each trade on the underlying's move over the same hold,
gate the residual) is now part of the bar for this family, the drift
decomposition is carried in the snapshot itself, and benchmark-relative
payoffs are available to every campaign.
\item[A one-direction discovery window.] Gates evaluated on one window cannot
see that the window had one direction. The rule recorded with the grid is
that a condition is read together with its opposite, and a passing pair is a
long-side artifact, not a signal.
\item[Costs assumed, not measured.] Two candidates were killed by a
doubled-cost gate against a cost nobody had measured, and one holdout was
spent before its cost premise was measured. The overnight SPX promotion was
spent under time pressure on a single night's cost capture and came back
rejected. Campaigns now report the breakeven cost alongside the result, and
the written rule is that measured entry costs precede any holdout spend.
\item[Look-ahead through a feature.] The single formal pass among 138 Kalshi
hypotheses conditioned on volume in the same window it predicted. Cells that
condition on in-window information are now marked descriptive-only and cannot
count as evidence.
\item[A lost specification.] See the holdout case above; the contract is now
persisted with the ledger.
\item[Spending evidence for nothing.] When every premise was already negative
in validation, the TA holdout was left unread: no outcome could change, so
reading it would only destroy a clean window.
\end{description}

\section{Limitations}

The loop makes it hard to report something that is not there; it does not make
true things easy to find. Its vocabulary is deliberately small, and a hypothesis
outside it cannot be tested until trusted code is extended. Several campaigns
use short samples (for example 46 days of Kalshi data, 60 weekends of ETF flow),
and their nulls are statements about power as much as about markets. Option
results from 2024 onward cover no bear market of 2022's kind. Some modeled
prices remain where actual quotes were not available; those campaigns say so.
Costs in several campaigns are assumptions with stated breakevens rather than
measurements. Finally, this paper is written by the system's author; the full campaign
records behind each number are available from the author on request.

\section{Conclusion}

A research loop in quantitative finance should be judged first by how hard it
is to fool and only second by what it finds. Making the model propose data
instead of code, separating stages physically, letting the holdout be spent
once, accepting only on a conjunction of absolute gates, refusing duplicates
before they cost a trial and keeping every failure in a chained ledger
together produce a system whose most common output is a well-documented
``no''. We think that is the correct base rate, and that publishing it is the
point.

\paragraph{Reproducibility.} Engine code, a reference mission and the
public pre-registration with its hash-chained ledger are at\\
\repo.\\
The test suite runs offline without keys. Campaign records are available from
the author on request.
A browser demonstration runs the engine under Pyodide on a synthetic tape with
a planted regime that discovery finds and validation refutes.

\aboutauthor

\paragraph{Disclaimer.} All backtests are hypothetical research evidence and
not investment advice.


\begin{thebibliography}{99}\small

\bibitem{zhang2026engine}
H.~Zhang.
\newblock Autonomous Quant Researcher.
\newblock \repo, 2026.

\bibitem{bailey2014deflated}
D.~H. Bailey and M.~L{\'o}pez de Prado.
\newblock The deflated Sharpe ratio: correcting for selection bias, backtest
overfitting, and non-normality.
\newblock \emph{Journal of Portfolio Management}, 40(5):94--107, 2014.

\bibitem{bailey2017probability}
D.~H. Bailey, J.~M. Borwein, M.~L{\'o}pez de Prado, and Q.~J. Zhu.
\newblock The probability of backtest overfitting.
\newblock \emph{Journal of Computational Finance}, 20(4):39--69, 2017.

\bibitem{bailey2014pseudo}
D.~H. Bailey, J.~M. Borwein, M.~L{\'o}pez de Prado, and Q.~J. Zhu.
\newblock Pseudo-mathematics and financial charlatanism: the effects of
backtest overfitting on out-of-sample performance.
\newblock \emph{Notices of the American Mathematical Society}, 61(5):458--471,
2014.

\bibitem{greyling2026loop}
C.~Greyling.
\newblock Loop Engineering.
\newblock \url{https://github.com/cobusgreyling/loop-engineering}, 2026.

\bibitem{hansen2005test}
P.~R. Hansen.
\newblock A test for superior predictive ability.
\newblock \emph{Journal of Business \& Economic Statistics}, 23(4):365--380,
2005.

\bibitem{harvey2016cross}
C.~R. Harvey, Y.~Liu, and H.~Zhu.
\newblock \ldots and the cross-section of expected returns.
\newblock \emph{Review of Financial Studies}, 29(1):5--68, 2016.

\bibitem{hou2020replicating}
K.~Hou, C.~Xue, and L.~Zhang.
\newblock Replicating anomalies.
\newblock \emph{Review of Financial Studies}, 33(5):2019--2133, 2020.

\bibitem{karpathy2026autoresearch}
A.~Karpathy.
\newblock autoresearch.
\newblock \url{https://github.com/karpathy/autoresearch}, 2026.

\bibitem{kunsch1989jackknife}
H.~R. K{\"u}nsch.
\newblock The jackknife and the bootstrap for general stationary observations.
\newblock \emph{Annals of Statistics}, 17(3):1217--1241, 1989.

\bibitem{li2025rdagentquant}
Y.~Li et~al.
\newblock R\&D-Agent-Quant: a multi-agent framework for data-centric factors
and model joint optimization.
\newblock arXiv:2505.15155, 2025.

\bibitem{lo1990data}
A.~W. Lo and A.~C. MacKinlay.
\newblock Data-snooping biases in tests of financial asset pricing models.
\newblock \emph{Review of Financial Studies}, 3(3):431--467, 1990.

\bibitem{mclean2016does}
R.~D. McLean and J.~Pontiff.
\newblock Does academic research destroy stock return predictability?
\newblock \emph{Journal of Finance}, 71(1):5--32, 2016.

\bibitem{nosek2018preregistration}
B.~A. Nosek, C.~R. Ebersole, A.~C. DeHaven, and D.~T. Mellor.
\newblock The preregistration revolution.
\newblock \emph{Proceedings of the National Academy of Sciences},
115(11):2600--2606, 2018.

\bibitem{tang2025alphaagent}
Z.~Tang et~al.
\newblock AlphaAgent: LLM-driven alpha mining with regularized exploration to
counteract alpha decay.
\newblock arXiv:2502.16789, 2025.

\bibitem{white2000reality}
H.~White.
\newblock A reality check for data snooping.
\newblock \emph{Econometrica}, 68(5):1097--1126, 2000.

\bibitem{xiao2024tradingagents}
Y.~Xiao, E.~Sun, D.~Luo, and W.~Wang.
\newblock TradingAgents: multi-agents LLM financial trading framework.
\newblock arXiv:2412.20138, 2024.

\bibitem{barton2026production}
T.~J. Barton, C.~Constantakis, P.~Hauseman, A.~Mous, A.~Hoffman, B.~Bergeron,
and H.~Goodreau.
\newblock What LLM trading agents actually do in production: a six-month,
population-scale record from two fleets.
\newblock arXiv:2609.05663, 2026.

\bibitem{chen2026nowcasting}
Z.~Chen and D.~Pu.
\newblock Autonomous market intelligence: agentic AI nowcasting predicts stock
returns.
\newblock arXiv:2601.11958, 2026.

\bibitem{gencay2026honest}
E.~Gen\c{c}ay.
\newblock What survives honest evaluation? Leakage-safe, search-aware
assessment of LLM-driven trading strategy discovery.
\newblock arXiv:2608.27734, 2026.

\bibitem{deep2025walkforward}
G.~Deep, A.~Deep, and W.~Lamptey.
\newblock Interpretable hypothesis-driven trading: a rigorous walk-forward
validation framework for market microstructure signals.
\newblock arXiv:2512.12924, 2025.

\bibitem{li2025timetravel}
C.~Li, Y.~Shi, et~al.
\newblock Time travel is cheating: going live with DeepFund for real-time fund
investment benchmarking.
\newblock arXiv:2505.11065, 2025.

\bibitem{liu2025cognitive}
F.~Liu, Y.~Huang, et~al.
\newblock Cognitive alpha mining via LLM-driven code-based evolution.
\newblock arXiv:2511.18850, 2025.

\bibitem{romano2005stepwise}
J.~P. Romano and M.~Wolf.
\newblock Stepwise multiple testing as formalized data snooping.
\newblock \emph{Econometrica}, 73(4):1237--1282, 2005.

\bibitem{shi2025alphajungle}
Y.~Shi, Y.~Duan, and J.~Li.
\newblock Navigating the alpha jungle: an LLM-powered MCTS framework for
formulaic factor mining.
\newblock arXiv:2505.11122, 2025.

\bibitem{yao2026execution}
J.~Yao, Z.~Zheng, and B.~Li.
\newblock Execution realism and reproducibility in LLM-based trading systems: a
systematic scoping review and evidence audit.
\newblock arXiv:2606.08285, 2026.

\bibitem{zhao2025alphaagents}
T.~Zhao, J.~Lyu, S.~Jones, H.~Garber, S.~Pasquali, and D.~Mehta.
\newblock AlphaAgents: large language model based multi-agents for equity
portfolio constructions.
\newblock arXiv:2508.11152, 2025.

\bibitem{choi2025texttoalpha}
C.~Choi, Y.~Kim, Y.~Yu, Y.~Cha, V.~Z. Golkhou, I.~Halperin, et~al.
\newblock From text to alpha: can LLMs track evolving signals in corporate
disclosures?
\newblock arXiv:2510.03195, 2025.

\bibitem{yu2023finmem}
Y.~Yu et~al.
\newblock FinMem: a performance-enhanced LLM trading agent with layered memory
and character design.
\newblock arXiv:2311.13743, 2023.

\end{thebibliography}
\end{document}